\documentclass[]{raa}
\usepackage{graphicx,times}
\usepackage{natbib}
\usepackage{amssymb,amsmath}
\bibpunct{(}{)}{;}{a}{}{,}

\usepackage[a4paper=true,dvipdfm=true,pagebackref=true]{hyperref}
\hypersetup{pdftitle = The title of my PDF, pdfauthor = My name, pdfsubject= The subject, pdfkeywords = keyword1 keyword2 keyword3}
\hypersetup{colorlinks = true, linkcolor = green, anchorcolor = red, citecolor = blue, filecolor = red, pagecolor = red, urlcolor = red}

\begin{document}

   \title{AGN lifetimes in UV-selected galaxies: a clue to supermassive black hole-host galaxy coevolution}

 \volnopage{ {\bf 20XX} Vol.\ {\bf X} No. {\bf XX}, 000--000}
   \setcounter{page}{1}

   \author{Xiaozhi Lin\inst{1,2}, Yongquan Xue\inst{1,2}, Guanwen Fang\inst{3}, Lulu Fan\inst{1,2,4}, Huynh Anh N. Le\inst{1,2,5}, and Ashraf Ayubinia\inst{1,2}
   %Tao Wang\inst{4}, Xu Kong\inst{1,2},
   }
%% Here is an example of three authors come from different institutes.
%% For single author or all the authors from an institute, use "\inst{}" only

   \institute{ CAS Key Laboratory for Research in Galaxies and Cosmology, Department of Astronomy, University of Science and Technology of China, Hefei 230026, China ; xzlin@mail.ustc.edu.cn, xuey@ustc.edu.cn\\
%% Please give the E-mail address of the author, to whom future correspondence and
%% offprint requests will be sent.
        \and
             School of Astronomy and Space Science, University of Science and Technology of China, Hefei 230026, China\\
	    \and
             School of Mathematics and Physics, Anqing Normal University,  Anqing 246133, China\\
        \and
             Shandong Provincial Key Lab of Optical Astronomy and Solar-Terrestrial Environment, Institute of Space Science, Shandong University, Weihai, 264209, China\\
        \and
             CAS President's International Fellowship Initiative (PIFI) Fellow\\
%	\and
%             School of Astronomy and Space Science, Nanjing University, Nanjing 210093, People's Republic of China\\
\vs \no
   {\small Received 20XX Month Day; accepted 20XX Month Day}
}

\abstract{The coevolution between supermassive black holes (SMBHs) and their host galaxies has been proposed for more than a decade, albeit with little direct evidence about black hole accretion activities regulating galaxy star formation at $z>1$.
In this paper, we study the lifetimes of X-ray active galactic nuclei (AGNs) in $UV$-selected red sequence (RS), blue cloud (BC) and green valley (GV) galaxies, finding that AGN accretion activities are most prominent in GV galaxies at $z\sim1.5-2$, compared with RS and BC galaxies.
We also compare AGN accretion timescales with typical color transition timescales of $UV$-selected galaxies.
We find that the lifetime of GV galaxies at $z\sim1.5-2$ is very close to the typical timescale when the AGNs residing in them stay in the high-accretion-rate mode at these redshifts;
for BC galaxies, the consistency between the color transition timescale and the black hole strong accretion lifetime is more likely to happen at lower redshifts ($z<1$).
Our results support the scenario where AGN accretion activities govern $UV$ color transitions of host galaxies, making galaxies and their central SMBHs coevolve with each other.
}

   \authorrunning{X.-Z. Lin et al. }            %author_head in even pages
   \titlerunning{AGN lifetimes in UV-selected galaxies}  % title_head in odd pages
   \maketitle

%________________________________________________ sections below
%
\section{Introduction}           %% first-level sections will be auto-capitalized
\label{sect:intro}

\hspace{5mm}The coevolution of supermassive black holes (SMBHs) and their host galaxies has been one of the most fundamental questions in galaxy physics in the past two decades, since the discovery of the tight relation between black hole mass ($M_{\rm BH}$) and stellar velocity dispersion ($\sigma$), i.e., $M_{\rm BH}-\sigma$ relation \citep[e.g.,][]{Ferrarese_2000,Gebhardt_2000,Kormendy_2011,Kormendy_2013,Saglia_2016}.
Over three orders of magnitude, the correlation between black hole mass and stellar velocity dispersion only has an scatter of $\sim0.3~{\rm dex}$.
The studies of luminosity functions (and/or stellar mass functions) have also revealed that the peaks of the global galaxy star formation and black hole accretion are both at $z\sim1.5-2$, and both show ``downsizing'' features, i.e., more massive galaxies/black holes started to exist and grow at earlier epoches of the cosmic time \citep[e.g.,][]{Miyaji_2000,Hopkins_2007,Croom_2009}.
Such evidence has shown that SMBHs and their host galaxies are very likely to coevolve with each other.
In this case, some internal feedback mechanisms may be at work to govern the evolution.

One of the most popular scenarios to explain the coevolution between SMBHs and their host galaxies is through active galactic nucleus (AGN) feedback.
The galaxy star formation and the central black hole accretion are fueled by the gas reservoirs within the same galaxy.
When the dark matter halo grows to some critical values \citep[e.g.,][]{Hickox_2009,Lin_2019,Lin_2021}, gas-rich mergers or disk instabilities \citep[e.g.,][]{Martig_2009,Leauthaud_2012,Zolotov_2015} may cause the gas clumps to lose their angular momenta, fall down to the galaxy center, and enhance the star formation activities, which would further trigger the AGN feedback activities, depleting the surrounding gas, suppressing the galaxy star formation, and making the stellar component and central SMBH grow synchronously \citep{Merloni_2010,Graham_2012,Kormendy_2013,Sun_2015,Saglia_2016}.

Nonetheless, there has been little direct observational evidence showing that AGN feedback would definitely quench galaxy star formation at high redshifts \citep{TorresPapaqui_2020}.
The reasons, on one hand, lie in the observational difficulties in galactic scales at these redshifts.
On the other hand, the timescales that AGNs stay active (in the Eddington or super-Eddington phase), usually being several tens of million years, are much shorter than the typical star formation timescales of galaxies (more than several hundred million years) \citep{Bell_2002,Martini_2004,Adelberger_2005,Hopkins_2005,Hopkins_2009a}.
Thus, it is difficult to observe the interactions between AGN feedback and galaxy global star formation activities in the early universe.

In galaxy physics, the color bimodality (i.e., blue vs. red) of the galaxy distribution has long been known effective in classifying and studying the star-forming and quiescent galaxies up to high redshifts \citep{Strateva_2001,Blanton_2003,Bell_2004,Giallongo_2005,Borch_2006,Franzetti_2007,Cirasuolo_2007,Cassata_2008,Whitaker_2011,Lee_2015}.
Although some dust-reddened star-forming galaxies may contaminate the red population, some researchers modified the separation criterion to improve the galaxy classification.
For example, \citet{Brammer_2009} introduced the dust-attenuation factor into the color separation criterion, making it more effective to classify the intrinsic star-forming and quiescent galaxies.
\citet{Wang_2017} checked out this new separation criterion, making it more self-consistent from the local to the high
redshift universe ($z\sim2.5$).

There is a subtle special population of galaxies lying in the color conjunction region between the red and blue bimodality, which are the so-called green valley (GV) galaxies \citep[e.g.,][]{Brammer_2009,Xue_2010,Wetzel_2012,Fang_2013,Schawinski_2014,Chang_2015}.
These galaxies usually show strong post-starburst features \citep[e.g.,][]{Rowlands_2018}, and are considered as experiencing the quenching of star formation.
Studying the relevant properties of these galaxies as well as their evolution with redshift hence can help to identify whether AGN feedback activity plays an important role during galaxy quenching.
Some studies have revealed that the fraction of galaxies hosting AGNs is higher in GV galaxies, compared to red sequence (RS) and blue cloud (BC) galaxies \citep[e.g.,][]{Silverman_2008,Schawinski_2010,Wang_2017}.
However, most of the evidence about the AGN activities interacting with the star formation within these galaxies is indirect \citep[e.g.,][]{Silk_1998,Ishibashi_2012,Zubovas_2013,Cresci_2015,Le_2017a,Le_2017b,Shin_2019}.
If AGN feedback indeed drives the quenching within these galaxies, the timescales that the central black holes remain active should be comparable to the color transition timescales of GV galaxies.

\citet{Lin_2021} measured the lifetimes of massive GV galaxies at high redshifts based on the clustering analyses.
The results can be rough estimates of the typical quenching timescales within these galaxies, shifting from the star-forming (BC) to quiescent (RS) phase, and are shown to be well consistent with the results from spectral energy distribution (SED) modeling (also see \citealt{Lin_2021}).
Meanwhile, measuring the lifetimes of AGNs within these galaxies is also possible.
\citet{Hopkins_2009a} quantified the quasar lifetimes through their Eddington ratio ($\lambda_{\rm Edd}$) distribution.
They demonstrated that, as long as the lifetimes of AGNs are short relative to the Hubble time, probing their lifetimes uniquely through the Eddington ratio distribution is applicable.
Therefore, if we know the distributions of Eddington ratio in RS, BC and GV galaxies, we can estimate the corresponding lifetimes of AGNs among these galaxies, and the results can be directly compared to the typical color transitional timescales of galaxies (i.e., lifetimes of galaxies of specific colors).
In this way, we can examine the roles that AGNs play in quenching galaxy star formation.

Fortunately, \citet{Wang_2017} provided the Eddington ratio distributions of AGNs in the galaxies of different extinction-corrected $UV$ colors through a systematic statistics of the fractions of AGNs of different accretion rates.
Based on their results, we can estimate the lifetimes of AGNs of different accretion rates in galaxies of different colors in each specific cosmic epoch.
Thus, it is possible to compare the AGN lifetimes with the typical color transitional timescales of galaxies and check the current quenching scenario, pinning down whether AGN feedback is important, especially in GV galaxies.

The outline of this paper is as follows. In Section~\ref{sect:method}, we introduce the method we adopt to compute the lifetimes of AGNs in RS, BC and GV galaxies. In Section~\ref{sect:lambda_edd}, we describe our adopted Eddington ratio distributions. In Section~\ref{sect:R&D}, we discuss our results and compare the lifetimes of AGNs with the typical color transitional timescales in the three galaxy populations. We summarize our results and conclusions in Section~\ref{sect:sum}.

Throughout this paper, we adopt a flat cosmology with $\Omega_{\rm M}=0.3$, $\Omega_{\Lambda}=0.7$, $H_0=70~{\rm km~s^{-1}~Mpc^{-1}}$, and a normalization of $\sigma_8=0.84$ for the matter power spectrum.

\section{Method to compute the lifetimes of AGNs}\label{sect:method}

\hspace{5mm}We compute the lifetimes of AGNs utilizing the method discussed in \citet{Hopkins_2009a}.
This method is based on the Eddington ratio distribution of AGNs.

In general, as long as the lifetimes of AGNs are short compared to the Hubble time at a given luminosity or Eddington ratio (i.e., cosmological evolution of triggering rates can be ignored in the redshift bin concerned), the duty cycle $\delta$ of a certain type of AGNs can be expressed not only by the fraction of active accretion time over the full time interval, but also by the fraction of number counts/density of such AGNs over the whole AGN population.
In this circumstance, we can translate the distribution of duty cycle at different Eddington ratios into the lifetimes of AGNs (also see Equation 3 in \citealt{Hopkins_2009a}):
\begin{equation}\label{equ:dtcycle_t}
\frac{d\delta}{d~\log\lambda}=\frac{1}{t_H(z)}\frac{dt}{d~\log\lambda}~~~~~(t\ll t_H),
\end{equation}
where $t_H(z)$ is the Hubble time at the given redshift, and $\lambda$ (or $\lambda_{\rm Edd}$) is the Eddington ratio defined as:
\begin{equation}
\lambda\equiv\frac{L_{\rm bol}}{L_{\rm Edd}}=\frac{L_{\rm bol}}{1.3\times10^{38}~erg~s^{-1}~(M_{\rm BH}/M_\odot)}.
\end{equation}
On the other hand, the duty cycle of AGNs can also be related to the black hole number density function $\Phi$ and Eddington ratio distribution $P(\lambda)$ at a given mass and accretion rate (also see Equation 4 in \citealt{Hopkins_2009a}):
\begin{equation}\label{equ:dtcycle_n}
\frac{d\delta}{d~\log\lambda}=\frac{dn(\lambda,M_{\rm BH})/n(M_{\rm BH})}{d~\log\lambda}=\frac{\Phi(\lambda|M_{\rm BH})}{n(M_{\rm BH})}=\frac{\Phi(\lambda|M_{\rm BH})}{\int_{-\infty}^{+\infty}\Phi(\lambda|M_{\rm BH})~d~\log\lambda}=\frac{dP(\lambda|M_{\rm BH})}{d~\log\lambda}
\end{equation}
Hence, the lifetimes of AGNs can be estimated using the combination of the above two equations.

Specifically, inverting Equation~\ref{equ:dtcycle_t}, we obtain:
\begin{equation}\label{equ:lft}
\frac{dt(\lambda|M_{\rm BH})}{d~\log\lambda}=t_H(z)\frac{d\delta}{d~\log\lambda}
\end{equation}
Integrating duty cycle $\delta$ above $\lambda$, we obtain $t(>\lambda)\equiv t_H(z)\delta(>\lambda)$.
This lifetime is directly determined by the observed Eddington ratio distribution $P(\lambda)$ at any redshift z, and is often referred to as the implied ``AGN lifetime'' in various observational statistics.
In physics, this lifetime represents the time that AGN systems spend on staying above the accretion rate $\lambda$ at a given redshift.

If the accretion triggering rate is relatively constant over the redshift range for AGNs of a given black hole mass, and the lifetime is short relative to the respective Hubble time, then the estimation through Equation~\ref{equ:dtcycle_t} and Equation~\ref{equ:dtcycle_n} is applicable.
Furthermore, the Eddington ratio distribution of the observed AGNs, independent of any other constraints, can be directly translated into their lifetime distribution.
Even if the accretion triggering rate strongly evolves with redshift and/or their characteristic lifetime is long compared to the Hubble time (e.g., low-redshift massive black holes, where their main growths had happened in much earlier epochs), utilizing the Eddington ratio distribution at a given redshift to predict the lifetime distribution of AGNs is still applicable. But, in this case, the actual duty cycle of an AGN is equivalent to their characteristic lifetime convolved with the distribution of accretion activity over the redshift bin of interest.
Even so, in practice, accounting for the accretion triggering distribution over the redshift bin when mapping the observed Eddington ratio distribution does not significantly alter the shape of the lifetime distribution.
In principle, this procedure only results in deriving a more accurate effective duty cycle than simply multiplying by $\sim1/t_H$.
More relevant detailed discussions can be referred to \citet{Yu_2004} and \citet{Hopkins_2005,Hopkins_2006,Hopkins_2006a}.
Thus, once an observed AGN luminosity function (of a given black hole mass) is well constrained at each luminosity and redshift bin of interest, the corresponding Eddington ratio distribution can be uniquely translated into the lifetime distribution.

This lifetime, reflecting the total timescale an AGN spends on some accretion rate levels, may show a variation in galaxies of different extinction-corrected $UV$ colors.
Comparing the AGN lifetimes in galaxies of different colors can give us the information about the duration of accretion activities of various levels in each redshift bin, which can help to examine the most recent AGN feedback scenario.
Besides, the lifetimes of AGNs in these galaxies of different colors can also serve as the contrasts of the typical color transitional timescales of the respective galaxy populations.
If it is indeed AGN feedback activity that drives the star-formation quenching, some similarities between the two timescales should be well identified.

At very high redshifts, the formula computed through the Eddington ratio distribution may under-predict the real lifetime of an AGN.
This is because the Hubble time $t_H$ around these redshifts is very short, and if we do not consider AGN triggering physics varying significantly with redshifts, the lifetime of an AGN here would become comparable to the Hubble time.
Thus, the condition in Equation~\ref{equ:dtcycle_t} that $t\ll t_H$ would no longer hold.
Nevertheless, this slightly under-estimated lifetime would still provide a meaningful comparison.
We would discuss this influence on the final results and conclusions later in Section~\ref{sect:R&D}.

In this study, we also compare the lifetimes of AGNs with the lifetimes (i.e., typical color transitional timescales) of galaxies of different extinction-corrected rest-frame $UV$ colors (see Section~\ref{sect:R&D_lft}).
These lifetimes of galaxies are also computed through their duty cycles.
But unlike the lifetimes of AGNs, the lifetimes of galaxies are derived based on their clustering results \citep{Lin_2019}.
The simple algorithm is as follows.
We assume that for each dark matter halo having a mass similar to the measured halo mass of one of the three galaxy populations, the galaxy residing in it would undergo the respective color phase on that galaxy population during its evolution over cosmic time.
According to this assumption, if we observe a galaxy whose halo mass is similar to one of the three galaxy populations but the galaxy does not exhibit the respective color, it means that this galaxy has not yet evolved into the phase of this galaxy population or it has already passed this phase.
Thus the lifetime of a galaxy population here can be a rough estimate of the time that this galaxy is alive in a respective phase of color.
Considering the quenching scenario, this galaxy lifetime can be also referred to as the ``typical color transitional timescales of galaxies''.
In \citet{Lin_2021}, we have demonstrated that these galaxy color transitional timescales are consistent with the SED modeling results assuming an exponentially declining star-formation history.
For more details, we refer the readers to \citet{Lin_2021}.

\section{Adopted Eddington ratio distributions}\label{sect:lambda_edd}

\hspace{5mm}In this paper, we estimate the lifetimes of AGNs based on the Eddington ratio distribution of black holes in RS, BC and GV galaxies in the combination of GOODS-North and GOODS-South fields provided by \citet{Wang_2017}.
They separated both AGN hosts and non-AGN hosts into RS, BC and GV galaxies according to their extinction-corrected rest-frame $UV$ colors.
The X-ray data of these AGNs have good exposures, which are comprised of the 2Ms main point-source catalog in the GOODS-North field \citep{Alexander_2003} and the 4Ms main point-source catalog in the GOODS-South field \citep{Xue_2011}.
Then, they evaluated the AGN Eddington ratio distributions in the three populations of galaxies based on the statistics of their AGN fractions.
During their statistics, they corrected for the observational incompleteness utilizing the $V_{\rm max}$ method, adopting the detection limits derived from the sensitivity maps in the GOODS-North \citep{Alexander_2003} and GOODS-South fields \citep{Xue_2011}.
Their detailed procedure is as follows.
In each specific bin of redshift and stellar mass, they computed the fraction of AGNs falling in a fixed narrow bin of $\lambda_{\rm Edd}$ around the median as the observed Eddington ratio distribution, $P(\lambda_{\rm Edd}|M_*,z)$.
Then, they performed a simple linear fit to each of the Eddington ratio distribution,
\begin{equation}
P(\lambda_{\rm Edd})~d~\log~\lambda_{\rm Edd}=A~(\frac{\lambda_{\rm Edd}}{\lambda_{\rm Edd_{\rm cut}}})^{-\alpha}~d~\log~\lambda_{\rm Edd},
\end{equation}
where A denotes the amplitude and $\alpha$ denotes the slope. $\lambda_{\rm Edd_{\rm cut}}$ is an arbitrary scaling factor, which was adopted as the Eddington luminosity, i.e., $\lambda_{\rm Edd_{\rm cut}}=1$.
This power-law form of fitting was also suggested and adopted in a number of similar relevant studies \citep[e.g.,][]{Aird_2012,Trump_2015,Jones_2016}.

\begin{figure}
   \centering
   \includegraphics[width=1.0\columnwidth, angle=0]{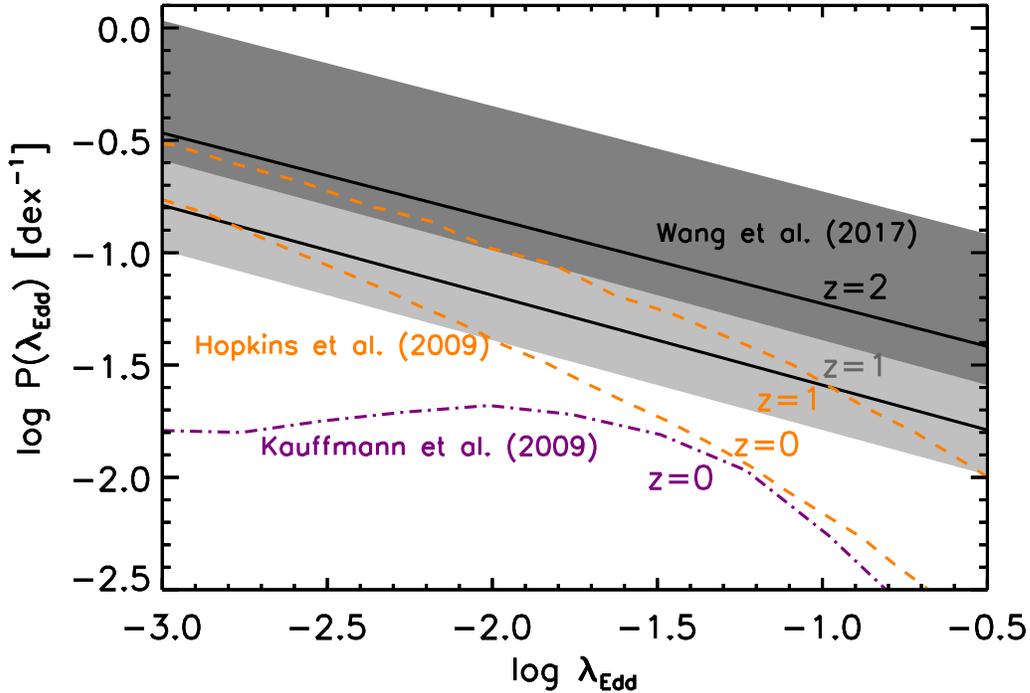}
  % \begin{minipage}[]{85mm}
   \caption{Eddington ratio distributions of the full AGN samples at $z=1$ and $z=2$ (black lines with dark and light grey regions) provided by \citet{Wang_2017} in comparison with the local relation presented by \citet{Kauffmann_2009} (purple dash-dotted line) and the semi-analytic model predictions in \citet{Hopkins_2009a} at $z=0$ and $z=1$ (yellow dashed lines).
Both the observational local empirical relation presented by \citet{Kauffmann_2009} and semi-analytic model predictions in \citet{Hopkins_2009a} are restricted to the SMBH masses of $10^7~M_\odot<M_{\rm BH}<10^8~M_\odot$.}
%\end{minipage}
   \label{fig:P_ER_all}
   \end{figure}

Figure~\ref{fig:P_ER_all} shows the Eddington ratio distributions of the full AGN samples at $z=1$ and $z=2$ (black lines with dark and light grey regions) provided by \citet{Wang_2017} in comparison with the local relation presented by \citet{Kauffmann_2009} (purple dash-dotted line) and the semi-analytic model predictions in \citet{Hopkins_2009a} at $z=0$ and $z=1$ (yellow dashed lines).
Observationally, \citet{Kauffmann_2009} measured the Eddington ratios of local AGNs based on the extinction-corrected [$\rm OIII$] emission line luminosity.
We adopted their Eddington ratio distribution selected with $10^7~M_\odot<M_{\rm BH}<10^8~M_\odot$ as analogues in the local universe, which corresponds to the galaxy stellar mass at $10^{10}~M_\odot<M_\ast<10^{11}~M_\odot$.
This stellar mass range is roughly identical to the galaxy samples in \citet{Wang_2017} and \citet{Lin_2021}.
As in \citet{Wang_2017}, we apply a $10\%$ normalization correction to the Eddington ratio distribution in \citet{Kauffmann_2009}, since AGN hosts only comprise $\sim10\%$ in the total amount of galaxies at the stellar mass of $M_\ast\sim10^{10.5}~M_\odot$ \citep{Trump_2015} and \citet{Kauffmann_2009} only presented the $\lambda_{\rm Edd}$ distribution in AGN host galaxies.
Apart from the result in \citet{Kauffmann_2009}, \citet{Hopkins_2009a} predicted the Eddington ratio distributions in different bins of black hole mass at $z=0$ and $z=1$.
Their results were based on the best-fit observational quasar luminosity function as a function of redshift \citep{Hopkins_2007}, as well as the best-fit semi-analytic lifetime distributions as a function of black hole mass in the hydrodynamic simulations of galaxy mergers incorporating with black hole growth \citep{Hopkins_2006}.
After integrating them over the redshifts of interest with $M_{\rm BH}$ fixed, they obtained the Eddington ratio distribution of quasars in each specific bin of $M_{\rm BH}$.
Same as the result quoted in \citet{Kauffmann_2009}, we adopt the $\lambda_{\rm Edd}$ distribution of AGNs in \citet{Hopkins_2009a} with a $M_{\rm BH}$ cut of $10^7~M_\odot<M_{\rm BH}<10^8~M_\odot$.

We note that utilizing the [$\rm OIII$] emission line flux as the indicator of the AGN accretion rate \citep{Kauffmann_2009} may not always be reliable, especially in the radiative inefficient regime (e.g., $L/L_{\rm Edd}<0.01$), where the ionized outflows and the star formation of the host galaxies may also produce low levels of [$\rm OIII$] line fluxes.
However, these influences are only restricted to the [$\rm OIII$] radiative inefficient cases, where the AGN samples also suffer some degrees of selection incompleteness.
And most importantly, in this paper, the AGN subsamples in \citet{Kauffmann_2009} are only served as the analogues of high-redshift AGNs in the local universe.
Therefore, we do not further consider the deviation of estimating Eddington ratio from [$\rm OIII$] emission line luminosity on the low accretion rate side.

In general, the Eddington ratio distributions of AGNs provided by \citet{Wang_2017} show a good consistency when compared to the observational statistical relations and/or semi-analytic model predictions at different redshifts.
Therefore, we do not consider any significant difference in the $\lambda_{\rm Edd}$ distribution arising due to the observational field variance in this paper.

\section{Results and Discussions}\label{sect:R&D}

\hspace{5mm}In this section, we present and discuss the main results of this paper.
In Section~\ref{sect:R&D_AR}, we compare the evolution of Eddington ratio distributions of AGNs as a function of redshift among galaxies of different extinction-corrected rest-frame $UV$ colors.
We also try to explain the possible physical mechanisms that are responsible for these trends.
In Section~\ref{sect:R&D_lft}, we display the AGN lifetimes among RS, GV and BC galaxies computed through the Eddington ratio distributions.
We study their redshift evolutions and compare the results with the typical color transitional lifetimes in the three population of galaxies.
In this way, we attempt to examine the possible roles of AGN accretion activities in the most current galaxy star-formation quenching scenario.

\subsection{Eddington ratio distributions of AGNs among RS, GV and BC galaxies across different redshifts}\label{sect:R&D_AR}

\begin{figure}
   \centering
   \includegraphics[width=1.0\columnwidth, angle=0]{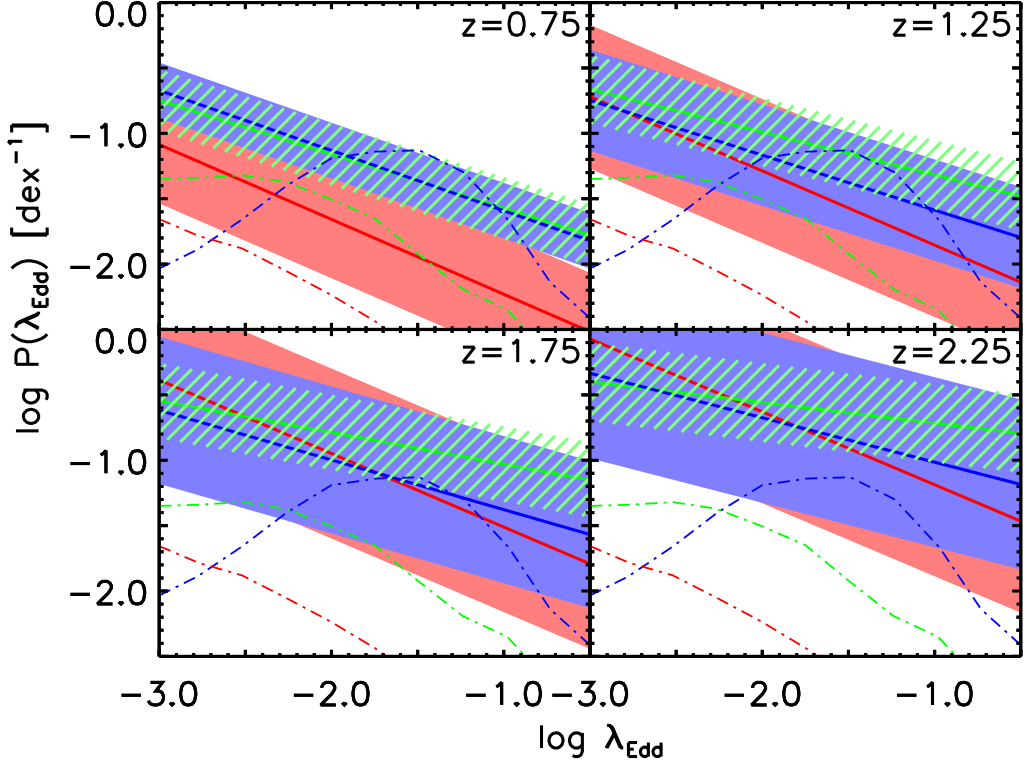}
  % \begin{minipage}[]{85mm}
   \caption{Eddington ratio distributions in BC, GV and RS galaxies at $z=0.75$, $z=1.25$, $z=1.75$, and $z=2.25$, shown as the thick solid color lines with the shaded/line-filled regions. The thin dash-dotted lines in each panel are the same relations of the respective local AGN host analogues from \citet{Kauffmann_2009}, selected with $1.2<D_n(4000)<1.4$, $1.6<D_n(4000)<1.7$ and $1.7<D_n(4000)<1.9$, respectively. Redder galaxies generally show a stronger redshift evolution of $\lambda_{\rm Edd}$ distribution than bluer ones.}
%\end{minipage}
   \label{fig:P_ER_sub}
   \end{figure}

\hspace{5mm}For each population of galaxies selected with different extinction-corrected rest-frame $UV$ colors, we compare the Eddington ratio distribution \citep{Wang_2017} with that of the local SDSS AGN host galaxies of different $4000~\AA$ break strength ($D_n(4000)$) in \citet{Kauffmann_2009}.
In particular, the AGN host galaxies selected with $1.2<D_n(4000)<1.4$, $1.6<D_n(4000)<1.7$ and $1.7<D_n(4000)<1.9$ are chosen as the typical analogues of the BC, GV and RS galaxies in \citet{Wang_2017}, respectively.
Considering that \citet{Wang_2017} only provided the $\lambda_{\rm Edd}$ distributions of galaxies at $z=1$ and $z=2$, we interpolate their best-fit amplitudes and slopes of the $\lambda_{\rm Edd}$ distribution functions of RS, GV and BC galaxies, respectively, as a function of redshift to predict the Eddington ratio distribution of galaxies of each color at four more specific redshifts, i.e., $z=0.75$, $z=1.25$, $z=1.75$, and $z=2.25$.
When accounting for the uncertainties of the Eddington ratio distribution, we keep the slope fixed for a clearer visualization.
The result is shown in Figure~\ref{fig:P_ER_sub}.

It is clear that the Eddington ratio distributions of galaxies selected with different extinction-corrected rest-frame $UV$ colors are closer to the local analogues of the respective $D_n(4000)$ indexes in \citet{Kauffmann_2009} towards lower redshifts.
The redshift evolution of the distribution of $\lambda_{\rm Edd}$ is much more significant in $UV$ redder galaxies.
The AGN accretion rate in low-redshift or local RS galaxies is very low.
These galaxies are usually considered as those having already consumed their gas supply, and both of their star formation and central black hole accretion tend to stop.
While at higher redshifts, the difference between the $\lambda_{\rm Edd}$ distributions of $UV$ red and blue galaxies becomes very small, especially around the medium and low accretion rates.
This phenomenon may reflect the possible situation that gas supply to the central black hole has not yet been completely shut off in high-redshift RS galaxies, thus their AGN accretion rates remain in a moderate level.
On the other hand, the accretion rates of central black holes among BC galaxies do not show significant evolution with redshifts, which indicates that there may not be significant difference in the black hole accretion triggering mechanisms among these galaxies.
BC galaxies are usually found to contain plenty of molecular gas supplies at whatever redshifts \citep[e.g.,][]{Stark_1986,Young_1989,Sage_1992,Taylor_1999,Barone_2000,Boselli_2014,Canameras_2018,Tacconi_2018}.
Hence, it is quite likely that the gas in these galaxies would freely assemble and fall down to feed the central black hole without any interruption.

There are a few possibilities that some dust-rich galaxies with moderate accretion rates would contaminate the RS galaxy population, especially at higher redshifts, i.e., z=1.75 and z=2.25.
But since \citet{Wang_2017} applied extinction-corrected rest-frame $UV$ colors to divide the galaxies, which is proven to be valid in separating the intrinsic red quiescent galaxies from the dust-reddened star-forming galaxies up to $z\sim2.5$, we thus do not consider this contamination effect in this paper.
It is also possible that some reverse mechanisms, such as star-formation rejuvenation, may compensate for the amount of galaxies lost during quenching.
But since in most studies, galaxies discovered to be experiencing rejuvenation only make up a tiny fraction of the total amounts \citep{Reipurth_1997,Portegies_2000,Perets_2009}, it is unlikely to significantly influence the global trends of galaxy quenching and the cosmic cooling.
We thus also do not take this effect into account in this paper.

\subsection{AGN lifetimes in galaxies of different extinction-corrected rest-frame $UV$ colors across different redshifts}\label{sect:R&D_lft}

\begin{figure}
   \centering
   \includegraphics[width=1.0\columnwidth, angle=0]{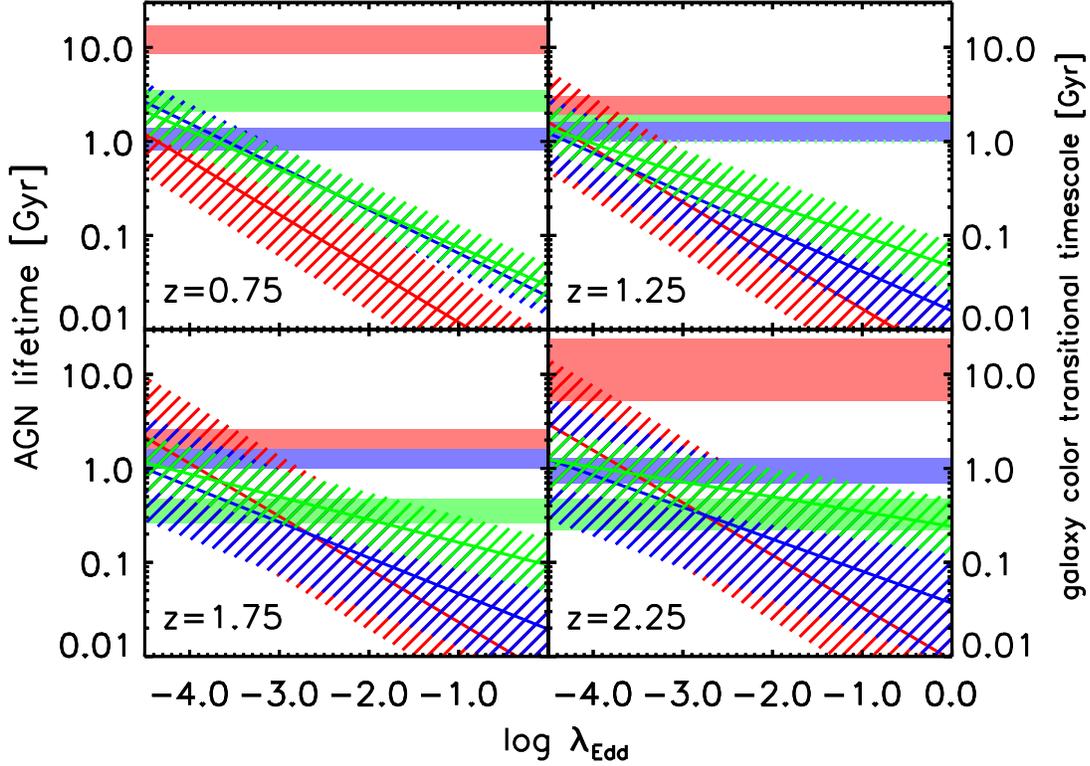}
  % \begin{minipage}[]{85mm}
   \caption{The lifetimes of AGNs among galaxies of different extinction-corrected rest-frame $UV$ colors at $z=0.75$, $z=1.25$, $z=1.75$, and $z=2.25$ in comparison with the typical color transitional timescales of these galaxies. Solid colored lines with the line-filled regions show the lifetimes and the corresponding uncertainties of AGNs among the RS, GV and BC galaxies derived from their Eddington ratio distributions following Equation~\ref{equ:dtcycle_t} and Equation~\ref{equ:dtcycle_n}. The horizontal color-filled regions show the typical color transitional timescales of the three galaxy populations at the corresponding redshifts taken from \citet{Lin_2021}.}
%\end{minipage}
   \label{fig:lft_AR}
   \end{figure}

\hspace{5mm}Based on the Eddington ratio distributions of AGNs among each of the above three populations of galaxies, we compute their respective lifetimes through Equation~\ref{equ:dtcycle_t} and Equation~\ref{equ:dtcycle_n}.
The result is shown in Figure~\ref{fig:lft_AR}.
In general, on the low accretion rate side, the lifetimes of AGNs among RS, BC and GV galaxies show no significant differences, especially at higher redshifts ($z>1$).
However, on the high accretion rate side, the lifetimes of AGNs in GV galaxies are obviously longer when compared to those of RS or BC galaxies, particularly at $z>1$.
The trend is even clearer at $z>1.5$.
These results indicate that strong AGN accretion is rather common in GV galaxies, in particular at higher redshifts ($z\sim2$).

Recently, there have been a number of studies revealing that the emergence of GV galaxies lying between the quiescent RS and the strong star-forming BC populations may be closely related to AGN activities.
For example,
local AGNs or changing-look quasars are usually preferentially found in GV galaxies \citep[e.g.,][]{Lacerda_2020,Liu_2021,Zhang_2021,Dodd_2021}.
The other possible connection between GV galaxies and AGNs is that both of the two populations may be experiencing ``morphological compaction'' \citep[e.g.,][]{Barro_2013,Barro_2015,Kocevski_2017,Lu_2021}.
But till now, there has been little direct evidence about AGN activities inevitably suppressing the global galaxy star formation, especially at high redshifts \citep[e.g.,][]{TorresPapaqui_2020}.
There are indeed some studies about AGN-driven star-formation quenching in the local universe.
For example, based on the Calar Alto Legacy Integral Field spectroscopy Area survey (CALIFA) integral-field unit (IFU) data, \citet{Kalinova_2021} traced the ionised hydrogen distribution maps in kiloparsec (kpc) scales to study galaxy quenching mechanisms.
Their results as well as some relevant studies \citep[e.g.,][]{Appleby_2020,Brownson_2020,Zewdie_2020} reinforced the ``inside-out'' quenching scenario where quenching begins from galaxy centers.
However, this evidence is still somewhat indirect, since when they examined the properties of those galaxies with weak and strong AGN emission features during their quenching processes, they do not find any observable differences.
The result of AGN lifetime estimation in galaxies of different extinction-corrected rest-frame $UV$ colors in this paper hence may complement the recent discoveries of AGN and host-galaxy coevolution, which provides us a clue to gain more information about AGN feedback at high redshifts.

To examine whether AGN accretion activities drive quenching in star-forming galaxies, we compare the AGN lifetime distribution in a certain population of galaxies with the respective typical galaxy color transitional timescale \citep{Lin_2021}.
In \citet{Lin_2021}, we adopted the same selection criteria as \citet{Wang_2017} to divide all COSMOS/UltraVISTA galaxies into populations of different extinction-corrected rest-frame $UV$ colors.
Then, we estimated the lifetime (i.e., color transitional timescale) of each galaxy population based on its derived duty cycle.
To achieve this, we first assumed that for each dark matter halo that has a similar mass with the measured host halo ($M_h$) of one of the galaxy populations in \citet{Lin_2019}, the galaxy residing in it would undergo the phase (color) of the respective galaxy population.
After that, we counted the total number of the halos that have similar $M_h$ to the host halo of that galaxy population in \citet{Lin_2019}, and computed the fraction of halos hosting those galaxies, i.e., the duty cycle.
The resulting lifetimes (i.e., color transitional timescales of the three galaxy populations) were shown to be well consistent with the results computed through SED modeling (for more details, we refer the readers to \citealt{Lin_2021}).

We show the typical galaxy color transitional timescales at each redshift as the horizontal color-filled areas in each panel of Figure~\ref{fig:lft_AR}.
It is clear that for GV galaxies, the region where the AGN lifetime and galaxy color transitional timescale overlap lies at low $\lambda_{\rm Edd}$ at lower redshifts, but moves to high $\lambda_{\rm Edd}$ at higher redshifts ($z\sim2$).
In other words, the typical color transitional timescales of GV galaxies coincide with the lifetimes of AGNs of different accretion activities at different redshifts.
At higher redshifts, the timescale that strong AGN accretion persists is almost identical with the color transitional timescale of GV galaxies, which implies that strong AGN feedback \citep[e.g., quasar/radiation-mode feedback,][]{Sijacki_2008,Fabian_2012,King_2015,Cielo_2018} is very likely to be accompanied with the color transition in GV galaxies.
At lower redshifts, the quenching timescale of GV galaxies becomes longer \citep[e.g.,][]{Schawinski_2014,Smethurst_2015,Belfiore_2017,Rowlands_2018,Trussler_2020}.
Given that the lifetimes of all AGNs are anti-correlated with the accretion rates, if the quenching (or galaxy color transition) here is still driven by AGN accretion activities, then weakly accreting AGNs should play the dominant role \citep[e.g.,][]{Shangguan_2018}.
This result can be reflected in Figure~\ref{fig:P_ER_sub}, where the slope of Eddington ratio distribution of GV galaxies becomes steeper at lower redshifts.

In comparison with GV galaxies, the typical color transitional timescale of BC galaxies always overlaps with their AGN lifetime distribution on the low $\lambda_{\rm Edd}$ side at whatever redshift.
We reckon that AGNs with high accretion rates would not have a quite significant effect on the color transition of the BC galaxy population.
This is because, although strong AGN feedback activities are likely to influence the star formation of the whole star-forming galaxy population, these extremely accreting AGNs only make up a small fraction of the total amount of AGNs within BC galaxies.
Besides, BC galaxies are usually discovered to be gas rich \citep{Stark_1986,Taylor_1999,Boselli_2014,Canameras_2018,Tacconi_2018}, which may effectively slow down the feedback wind driven by central AGNs.

We need to clarify that, although the inflowing gas within BC galaxies will feed the central SMBHs at a longer time than GV galaxies, which may further trigger subsequential strong AGN feedback, it does not mean that all the AGNs among BC galaxies are strongly accreting AGNs at the time we observed them.
\citet{Hickox_2009} proposed that the behaviors of AGNs and their host galaxies are accompanied by the growth of their host dark matter halos.
This scenario has been supported by a number of studies, \citep[e.g.,][]{Behroozi_2013,Cai_2013,Moster_2013,Vogelsberger_2014,Chen_2020}.
According to this scenario, the strong accretion and feedback activity of an AGN can only be triggered when the mass of its host halo reaches a critical value of $10^{12}\sim10^{13}~M_\odot$ (see Figure 16 in \citealt{Hickox_2009}).
For most BC galaxies, their typical halo masses have just reached the lower limit of this value \citep{Lin_2019}.
They also have relatively smaller stellar masses \citep{Lin_2019}, and thus smaller black hole masses compared with GV galaxies (if we assume a uniform black hole mass to stellar mass ratio).
Thus it is more likely that these less massive black holes among BC galaxies are accreting in a smooth and peaceful way.
They spend their long lifetimes in a relatively weaker accretion mode, as proposed in some studies \citep[e.g.,][]{Lapi_2014} that strong near Eddington accretion cannot persist long, and most of such accretions are more likely to happen in more massive SMBH systems.
Therefore, although some strongly accreting AGNs may act on the star formation in some specific BC galaxies, they are not able to cause the entire BC galaxy population to transform its color in a short time.
For the population of RS galaxies, we consider them as a stable phase according to the quenching scenario, so that their color transitional timescales are theoretically infinite, which can be much longer than the estimates shown in Figure~\ref{fig:lft_AR}.

\begin{figure}
   \centering
   \includegraphics[width=1.0\columnwidth, angle=0]{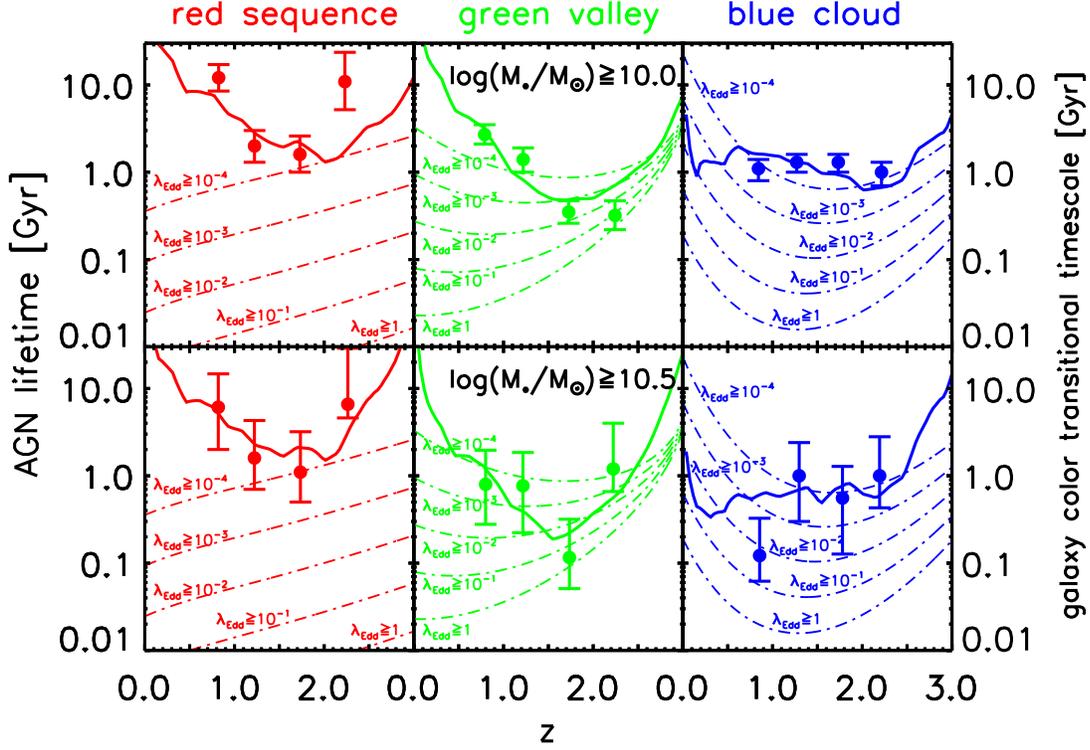}
  % \begin{minipage}[]{85mm}
   \caption{The comparison between the lifetimes of AGNs and the typical color transitional timescales of RS, BC and GV galaxies as a function of redshift. Upper panels show the results of host galaxies selected with $\log(M_\ast/M_\odot)\ge10.0$, while lower panels show the results of host galaxies selected with $\log(M_\ast/M_\odot)\ge10.5$. Filled circles with error bars in each panel show the lifetimes (typical color transitional timescales) of galaxies of different colors \citep{Lin_2021}. Thick solid curves show the best-fit lifetime parameters as a function of redshift to reproduce the exact photometric-redshift distributions of the three galaxy populations, based on a pre-assumed halo mass function \citep{Tinker_2008}, and the median halo mass growth rate from the Millennium-II simulation (see \citealt{Lin_2021} for more details). The thin dash-dotted lines in each panel show the lifetimes of the AGNs in the three galaxy populations, with the accretion rates fixed to $\lambda_{\rm Edd}\ge1,10^{-1},10^{-2},10^{-3},10^{-4}$. The lifetimes of the AGNs with high $\lambda_{\rm Edd}$ overlap with the typical color transitional timescales of GV galaxies at high redshifts ($z\sim1.5-2$).
   While among BC galaxies, the time when the lifetime curves of the high $\lambda_{\rm Edd}$ AGNs reach the typical color transitional timescales of BC galaxies seems to come much later, i.e., at $z<1$. It means that strong AGN feedback is very likely to be accompanied with the color transition in GV galaxies at high redshifts, reinforcing the coevolution scenario between central AGNs and host galaxies. In comparison, among BC galaxies, the significant star-formation regulation due to the strong central AGN accretion activities seems to be put off.
}
%\end{minipage}
   \label{fig:lft_z}
   \end{figure}

An alternative demonstration is to compare the lifetimes of AGNs with the galaxy color transitional timescales as a function of redshift, which is shown in Figure~\ref{fig:lft_z}.
Filled circles with error bars show the typical color transitional timescales of the three galaxy populations at $z\sim0.8,1.2,1.7$, and 2.2.
Thick solid curves show the best-fit color transitional lifetime parameters to reproduce the exact photometric-redshift distributions of the three galaxy populations, based on a pre-assumed halo mass function \citep{Tinker_2008} and a median halo mass growth rate in the Millennium-II simulations \citep{Fakhouri_2010}.
The typical lifetimes of AGNs with the accretion rates fixed to $\lambda_{\rm Edd}\ge1,0.1,0.01,0.001,0.0001$ for each galaxy population are shown as the dash-dotted lines of each color in the same panel.
These lifetimes are derived in a similar way as computing the AGN lifetimes as a function of accretion rate (see Figure~\ref{fig:lft_AR}).
The only difference is that here, we interpolate the AGN lifetime as a function of redshift rather than accretion rate.
During each interpolation, we allow redshift to vary and keep the accretion rate limit $\lambda_{\rm Edd}$ fixed.

A notable feature in Figure~\ref{fig:lft_z} is that the lifetimes of AGNs in galaxies of different colors exhibit distinct redshift evolution behaviors.
Among RS galaxies, for any given thresholds of their accretion rates, the lifetimes of AGNs increase smoothly with redshift.
The situations are different in GV and BC galaxies.
Among GV galaxies, while the lifetimes of the strongestly accreting AGNs still increase monotonically with redshift, those with medium and low accretion rates appear to have minimum lifetimes at $z\sim1-2$.
This trend is even more significant among BC galaxies, where the AGN lifetimes of all accretion rates reach the minimums at $z\sim1-2$.
These results indicate that observable strong and persistent AGN accretion activities are most likely to be detected among the high-redshift ($z\sim2$) GV galaxies and some of the lower-redshift BC galaxies.

Further comparison between the AGN lifetimes and the typical color transitional timescales of galaxies at each redshift reveals that the lifetimes of the strongly accreting AGNs among GV galaxies reach the typical color transitional timescales of GV galaxies at $z\sim1.5-2$, which is considered as the peak of the global black hole accretion activities and the peak of the global star-formation activities across cosmic time \citep{Delvecchio_2014,Heckman_2014,Madau_2014,Ma_2015,Sijacki_2015,Zakamska_2016}.
The result is consistent with the scenario where strongly accreting AGNs act on governing the gas depletion in GV galaxies at high redshifts, transforming the galaxy colors, making SMBHs and galaxies coevolve with each other during the cosmic time.
As a comparison, in BC galaxies, the time that strong AGN accretions significantly act on galaxy color transition comes much later ($z<1$).
The trend is clearer in more massive BC galaxies, i.e., $\log(M_\ast/M_\odot)\ge10.5$, which is shown in the lower-right panel of Figure~\ref{fig:lft_z}.
This result agrees well with the studies of \citet{Goncalves_2012} and \citet{Fritz_2014}, who found a rapid acceleration of quenching among BC galaxies at $z<1$, through analyzing the evolution of their luminosity/stellar mass functions.

Our study of AGN lifetimes is based on their Eddington ratio distributions.
As mentioned in Section~\ref{sect:method}, the condition of $t\ll t_H$ in Equation~\ref{equ:dtcycle_t} may not be established in the early universe, which would probably result in the under-prediction of the true lifetimes of the AGNs at high redshifts.
Nevertheless, this effect would not alter the main results and conclusions of this paper, since it only causes the lifetime of AGNs to rise at the high-redshift end in Figure~\ref{fig:lft_z}, which makes the lifetimes of the strongestly accreting AGNs better match the typical color transitional timescales of the GV galaxies at these redshifts.

At lower redshifts, however, it is also possible that the feedback activities of the central black holes may not be the only and main cause of galaxy quenching, since AGNs among galaxy populations of all colors spend their long lifetimes on the weakly accreting phase (see Figure~\ref{fig:lft_AR}).
Various hydrodynamical simulations considering AGN-driven quenching also tend to over-produce the quiescent GV galaxies (i.e., GV galaxies with lower specific star-formation rates) compared with the observations \citep{Angthopo_2021}, which suggests that there may be some additional mechanisms that dominate quenching at low redshifts, e.g., environmental quenching \citep{Schawinski_2014,Belfiore_2017,Kacprzak_2021}, halo quenching \citep{Bluck_2020}, or a modified more complex AGN-driven quenching mechanism such as halo mass threshold and black hole seed mass triggering AGN feedback \citep{Dave_2020,Trussler_2020,Angthopo_2021}.

\section{Summary and Conclusions}\label{sect:sum}

\hspace{5mm}In this paper, we follow the script proposed by \citet{Hopkins_2009a} to study the lifetimes of AGNs in the galaxies of different extinction-corrected rest-frame $UV$ colors, based on their derived Eddington ratio distributions in \citet{Wang_2017}.
First, we compare the Eddington ratio distribution of the entire galaxy population in \citet{Wang_2017} with the semi-analytic and observational results in \citet{Hopkins_2009a} and \citet{Kauffmann_2009} at lower redshifts, (i.e., $z=1$ and $z=0$).
We find that they are well consistent with each other.

Second, we make a systematic comparison between the Eddington ratio distributions of the AGNs among the galaxies of different $UV$ colors.
We also compare them to the $\lambda_{\rm Edd}$ distributions of the AGNs among local galaxies of several respective $D_n(4000)$ indexes \citep{Kauffmann_2009} to study the evolution of $\lambda_{\rm Edd}$ as a function of redshift. The $\lambda_{\rm Edd}$ distribution of the AGNs among RS galaxies exhibits the strongest evolution among the three galaxy populations; while for BC galaxies, we do not detect any significant evolution of the $\lambda_{\rm Edd}$ distribution with cosmic time, indicating that there may be no apparent difference in the black hole accretion triggering mechanisms among these galaxies.

Third, we study the lifetimes of the X-ray AGNs among the galaxies of different extinction-corrected rest-frame $UV$ colors.
The typical lifetimes of X-ray AGNs among these galaxies are generally from several tens of million years to $\sim1~{\rm Gyr}$, with strongerly accreting sources having shorter lifetimes.
On the low accretion rate side, there are no observable differences in the estimated lifetimes among the galaxies of all colors.
However, on the high accretion rate side, it is clear that the AGNs among GV galaxies have the longest lifetimes compared to those of RS and BC galaxies, suggesting that strongly accreting AGNs are rather common in GV galaxies, especially at higher redshifts ($z\sim2$).

Fourth, we compare the lifetimes of the AGNs with the typical color transitional timescales of the galaxies of all colors derived in \citet{Lin_2021} as a function of $\lambda_{\rm Edd}$.
The typical color transitional timescales of GV galaxies are consistent with the lifetimes of the weakly and strongly accreting AGNs at lower and higher redshifts, respectively, indicating an enhancement of AGN feedback activities among these galaxies at higher redshifts.
However, for BC galaxies, their typical color transitional timescales always overlap with the lifetimes of lower accretion rate AGNs, which means that strong AGN feedback activities among these galaxies, accounting for a small fraction, are not able to change their global in-situ colors within a short period of time.

Finally, we study the AGN lifetimes and compare them to the typical color transitional timescales of galaxies as a function of redshift.
The lifetimes of the AGNs among the galaxies of different colors exhibit distinct redshift evolution behaviors.
It seems clear that strong and persistent AGN accretion is most detectable among those high-redshift ($z\sim2$) GV galaxies and some of the lower-redshift ($z<1$) BC galaxies.
The lifetimes of the strongly accreting AGNs among GV galaxies are almost identical with the typical color transitional timescales of their hosts at $z\sim1.5-2$.
This redshift coincides with the peaks of global black hole accretion activities and galaxy star-formation activities during the entire cosmic time.
While in BC galaxies, strong AGN accretion activities are likely to have observable effects on the color transition of their hosts at lower redshifts ($z<1$).
Our results support the quenching scenario in massive galaxies where strong AGN accretion activities help to remove the gas contents in some intermediate-redshift star-forming galaxies, making SMBHs and their hosts coevolve with each other.

\normalem
\begin{acknowledgements}

%We acknowledge the constructive work by Wang Tao for presenting the statistical Eddington ratio distributions of AGNs in galaxies of different extinction-corrected rest-frame $UV$ colors.
%We acknowledge Philip F. Hopkins for the constructive work in proposing the script to compute the lifetime of AGNs based on their Eddington ratio distributions, as well as their semi-analytic work in deriving the Eddington ratio distributions of QSOs of different $M_{\rm BH}$ at $z\la1$.
%We acknowledge the hard work of Guinevere Kauffmann for presenting the Eddington ratio distributions of AGNs in the local universe based on the $O[III]$ emission line fluxes.
%This work is supported by the National Natural Science Foundation of China (Nos. 11673004, 23002601, 1320101002, 11433005, 12025303, 11890693, 11822303, 11773020, 12003031 and 11421303).
We thank the referee for helpful comments.
X.Z.L., Y.Q.X., and H.A.N.L. acknowledge support from NSFC-12025303, 11890693, 11421303, 12003031, the CAS Frontier Science Key Research Program (QYZDJ-SSW-SLH006), the K.C. Wong Education Foundation, and the science research grants from the China Manned Space Project with NO. CMS-CSST-2021-A06.
H.A.N.L. acknowledges support from the Chinese Academy of Sciences President's International Fellowship Initiative (Grant No. 2019PM0020).
%G.W.F. acknowledges the support from Yunnan young and middle-aged academic and technical leaders reserve talent program (No. 201905C160039), Yunnan ten thousand talent program-young top-notch talent and Yunnan Applied Basic Research Projects (2019FB007).

\end{acknowledgements}

\bibliographystyle{raa}
%\bibliography{bibtex}
%\bibliography{Reference}
\bibliography{ms.bbl}

\end{document}